\documentclass[twocolumn]{aa} 

\usepackage{graphicx}
\usepackage{color}
\usepackage{xcolor}
\usepackage{txfonts}
\newcommand{\Mpc}{\mathrm{~km~s^{-1}~Mpc^{-1}}}
\newcommand{\LCDM}{\rm{\Lambda}CDM}

\def\({\left(}
\def\){\right)}
\def\[{\left[}
\def\]{\right]}

\begin{document}

   \title{Revising the Hubble constant, spatial curvature and dark energy dynamics with the latest observations of quasars}

   \author{Tonghua Liu\inst{1,2}
          \and
          Shuo Cao\inst{2,3}\thanks{\email{caoshuo@bnu.edu.cn}}
          \and
           Xiaolei Li\inst{4}\thanks{\email{lixiaolei@hebtu.edu.cn}}
          \and
          Hao Zheng\inst{5}
          \and
          Yuting Liu\inst{2,3}
          \and
          Wuzheng Guo\inst{2,3}
          \and
          Chenfa Zheng\inst{2,3}
          }

   \institute{School of Physics and Optoelectronic, Yangtze University, Jingzhou 434023, China; \and Institute for Frontiers in Astronomy and Astrophysics, Beijing Normal University, Beijing 102206, China;
   \and
   Department of Astronomy, Beijing Normal University, Beijing 100875, China;
         \and College of Physics,
Hebei Normal University, Shijiazhuang 050024, China; \and School of Mechatronic Engineering, Lishui Vocational $\&$ Technical College, Lishui 323000, China }

   \authorrunning{Liu et al.}
   \titlerunning{Revising the Hubble constant and spatial curvature with quasars}

   \date{Received XXX/ Accepted XXX}

  \abstract
{In this paper, we use a newly compiled sample of ultra-compact structure in radio quasars and strong gravitational lensing systems with quasars acting as background sources to constrain six spatially flat and non-flat cosmological models ($\Lambda$CDM, PEDE and DGP). These two sets of quasar data (the time-delay measurements of six strong lensing systems and 120 intermediate-luminosity quasars calibrated as standard rulers) could break the degeneracy between cosmological parameters ($H_0$, $\Omega_m$ and $\Omega_k$) and therefore provide more stringent cosmological constraints for the six cosmological models we study. A joint analysis of the quasar sample provides model-independent estimations of the Hubble constant $H_0$, which is strongly consistent with that derived from the local distance ladder by SH0ES collaboration in the $\Lambda$CDM and PEDE model. However, in the framework of a DGP cosmology (especially for the flat universe), the measured Hubble constant is in good agreement with that derived from the the recent Planck 2018 results. In addition, our results show that zero spatial curvature is supported by the current lensed and unlensed quasar observations and there is no significant deviation from a flat universe. For most of cosmological model we study (the flat $\Lambda$CDM, non-flat $\Lambda$CDM, flat PEDE, and non-flat PEDE models), the derived matter density parameter is completely consistent with $\Omega_m\sim 0.30$ in all the data sets, as expected by the latest cosmological observations. Finally, according to the the statistical criteria DIC, although the joint constraints provide substantial observational support to the flat PEDE model, they do not rule out dark energy being a cosmological constant and non-flat spatial hypersurfaces.}


\keywords{Cosmological parameters -- Gravitational lensing: strong -- Quasars: general}

\maketitle

\section{Introduction}

The $\LCDM$ model, i.e., the cosmological constant plus cold dark matter model is so far the simplest and most natural model that could fit the observations of type Ia supernovae (SNe Ia) \citep{Riess07,Alam17}, Cosmic Microwave Background Radiation (CMB) \citep{Ade16}, and strong gravitational lensing \citep{cao2012SL,cao2014cosmic,Cao:2015qja}. However, it is still puzzled by several problems such as the coincidence problem and the fine-tuning problem \citep{cao2010testing}. Meanwhile, there exists some huge observational discrepancies if one tries to motivate $\Lambda$ as a zero-point quantum vacuum energy, regarding the estimations of different cosmological parameters in the framework of $\LCDM$ model. One of the major issues is the inconsistency between the Hubble constant ($H_0$) inferred from a
$\Lambda$CDM fit to the CMB data (temperature and polarization) from \textit{Planck} satellite \citep{Aghanim:2018eyx} and the local direct $H_0$ measurement by using so-called standard candles (SNe Ia and Cepheid variables) from the \textit{Supernova $H_0$ for the Equation of State} collaboration (SH0ES) \citep{Riess:2019cxk}. Such tension now has reached 4$\sigma$-6$\sigma$ with the accumulation of precise astrophysical observations. Since there is no evidence for considerable systematic errors in the Planck observation and the local measurements \citep{Riess:2019cxk,DiValentino:2018zjj,Follin:2017ljs,Leandros}, increasing attention is focusing on alternative cosmological models beyond $\LCDM$, e.g., the Early Dark Energy models \citep{Tanvi} and interacting dark energy models considering the interaction between dark energy and dark matter \citep{Gabriela,Valiviita,ZXG,Li17}. Some recent works suggested that the tension between the CMB and local determinations of the Hubble constant could be greatly reduced, within the generalized Chaplygin gas model \citep{Yang}. In the recent of \citet{Liam19}, a Phenomenologically Emergent Dark Energy (PEDE) model was introduced to shift the constraints on $H_0$, which demonstrated its potential in addressing the Hubble constant problem \citep{Liam19}. In addition, modifications to General Relativity (GR) theory \citep{Cruz,Hashim1,Hashim2,Briffa,Ren} or the well-known Dvali-Gabadadze-Porrati (DGP) model physically motivated by possible multidimensionality in the brane theory provides another way to deal with the cosmological constant problem and alleviate the Hubble tension \citep{Cao:2017ivt,Xu10,Xu14,Giannantonio08}. Note that all of these models could not only explain the late-time cosmic acceleration from different mechanisms, but also describe the large-scale structure distribution of the Universe (see \citet{Koyama16} for recent reviews). In this paper, we will explore the validity of three spatially flat and non-flat cosmological models ($\Lambda$CDM, PEDE and DGP), focusing on the time delay measurements from strongly lensed quasars and the angular size measurements of ultra-compact structure in radio quasars. Specially, the Hubble constant, spatial curvature and dark energy dynamics will be revisited with with such a newly compiled quasar sample.

It is well known that the time-delays from strong gravitational lensing systems provide an independent method to measure the Hubble constant (time-delays are inversely proportional to the $H_0$). For a specific strong gravitational lensing system, a distant active galactic nucleus (AGN), which usually acts as the background source, is gravitationally lensed into multiple images by a foreground early-type galaxy. Meanwhile, the light emitted from background sources at the same time will arrive at the Earth at different time. Due to the variable nature of quasars, the precise measurements of time-delays between multiple images are realizable by monitoring flux variations of the lens. Actually, the time-delays are directly related to the so-called "time-delay distance", a combination of three angular diameter distances
of the lensed quasar systems (from observer to lens, from observer to source, and from lens to source). Such idea was recently realized by the $H_0$ Lenses in COSMOGRAIL's Wellspring (H0LiCOW) collaboration \citep{Wong:2019kwg}, which presented the fits on the Hubble constant and other cosmological parameters using a joint analysis of six gravitationally lensed quasars. In the framework of six different cosmological models, their results showed that the derived Hubble constant is in agreement with local distance ladder measurement in the spatially flat $\Lambda$CDM model. However, the measured time-delays from lensed quasars, which is only primarily sensitive to $H_0$, whereas demonstrate their relatively weak constraints on other cosmological parameters. For instance, the determined value of the matter density parameter in the flat $\Lambda$CDM model, $\Omega_m=0.30^{+0.13}_{-0.13}$, would shift to $\Omega_m=0.24^{+0.16}_{-0.13}$ in the non-flat $\Lambda$CDM model. Moreover, the drawback of this method is that the fits on the Hubble constant are strongly model dependent, i.e. the value of $H_0$ would shift to $H_0=81.6^{+4.9}_{?5.3} \Mpc$ when the dynamics of dark energy is taken into consideration \citep{Wong:2019kwg}. For the discussions about model-independent measurements of $H_0$, we refer the reader to see the following works \citep{Liao19a,Liao20a,Lyu2020,Collett:2019hrr,Wei:2020suh,Qi2020arx,Tau19}.

On the other hand, the angular-size/redshift relation of the ultra-compact structures in unlensed radio quasars was also proposed for cosmological applications \citep{Kellermann93}. In the subsequent analysis, based on the milliarcsecond angular size measurements from very-long-baseline interferometry (VLBI) technique, \citet{Cao:2017ivt} demonstrated the possibility of using
intermediate-luminosity quasars as standard rulers for cosmological inference, covering the redshift range of $0.462<z< 2.73$. In the framework of such a reliable cosmological standard ruler extended to higher redshifts, great efforts have been made in the recent studies to set observational limits on different cosmologies \citep{Li17,ZXG}, which shows that radio quasars could provide quite stringent constraints on cosmological parameters. However, one issue which should be discussed is the strong degeneracy between the Hubble constant $H_0$, the matter density parameter $\Omega_m$ \citep{Cao:2017ivt}, the cosmic curvature $\Omega_k$ \citep{Qi:2018aio}, and the equation of state of dark energy $\omega$ \citep{DiValentino:2019qzk,DiValentino:2020hov,2019arXiv190809139H}. Therefore, one may expect that the combination of  the latest observations of quasars, i.e., the angular size of compact structure in radio quasars as standard rulers and the time delays from gravitationally lensed quasars, would break the degeneracy between the Hubble constant and other cosmological parameters, in the framework of different cosmological models of interest.
This paper is organized as follows: In Sec.~\ref{sec:models}, we summarize the cosmological models to be analyzed. In Sec.~\ref{sec:data}
we briefly describe the quasar data and the corresponding analysis method. In Sec.~\ref{sec:res} we report the results of constraints on
the Hubble constant, spatial curvature and dark energy dynamics with the latest quasar data. Finally, we give our discussion and conclusions in Sec.~\ref{sec:conclusion}.

\section{Cosmological models} \label{sec:models}

Let us now describe the models we are going to analyse in the next section with the data sets. In this paper, we concentrate on three classes of cosmological models in a spatially non-flat and flat universe, including the standard $\Lambda$CDM model, the Phenomenologically Emergent Dark Energy (PEDE) model, and the Dvali-Gabadadze-Porrati (DGP) model.

Assuming the Friedmann-Lematre-Robertson-Walker (FLRW) metric, with the non-flat universe filled with ordinary pressureless matter (cold dark matter plus baryons), dark energy, and negligible radiation, the Friedmann equation reads as
\begin{eqnarray}
H^2(z)=H_0^2[\Omega_m(1+z)^3+{\widetilde{\Omega}_{DE}}(z)
+\Omega_k(1+z)^2],
\end{eqnarray}
where $\Omega_{m}$ and $\Omega_{k}$ are the present values of the density parameters of dust matter and spatial curvature, respectively. The dark energy component of $\widetilde{\Omega}_{DE}(z)$ takes the following form of
\begin{equation}
\widetilde{\Omega}_{DE}(z)=\Omega_{DE}\exp\big\{3\int^z_0\frac{1+\omega(z')}{1+z'}dz'\big\},
\end{equation}
where $\Omega_{DE}$ is the present value of the dark energy density parameter, and the equation of state of dark energy is defined as $\omega(z)=p_{DE}/\rho_{DE}$, with $p_{DE}$ and $\rho_{DE}$ the pressure and energy density of dark energy, respectively. When $\omega(z)=-1$ this so-called XCDM parameterization reduces to the concordance $\Lambda$CDM model, with  $\widetilde{\Omega}_{DE}(z)=1-\Omega_{m}-\Omega_{k}$. Recently, another kind of Phenomenologically Emergent Dark Energy (PEDE) model proposed in \citet{Liam19,Liam20} has attracted a lot of attention. In this cosmological scenario, the density of dark energy, which has no effective presence in the past and emerges in the later times, is written as
\begin{equation}
\widetilde{\Omega}
_{DE}(z)=\Omega_{DE}\times[1-\tanh(\log_{10}(1+z))].
\end{equation}
Note that compared with $\Lambda$CDM cosmology, the PEDE model has no extra degree of freedom. For the third scenario, our idea of modifying the gravity is based on the assumption that our universe is embedded in a higher dimensional space-time, arising from the braneworld theory \citep{Dvali00a}. In the DGP model, the cosmic acceleration is reproduced by the leak of gravity into the bulk at large scales, which
result in the accelerated expansion of the Universe without the need of dark energy \citep{Dvali00b}. In the framework of a non-flat DGP model, the Friedman equation is modified as \citep{Deffayet02a,Deffayet02b,Tuomas03}
\begin{eqnarray}
H^2(z)=H_0^2[(\sqrt{\Omega_m(1+z)^3+\Omega_{r_c}}+\sqrt{\Omega_{r_c}})^2
+\Omega_k(1+z)^2],
\end{eqnarray}
where the density parameter $\Omega_{r_c}=1/(4r_c^2H_0^2)$ is associated with the length of $r_c$ where the leaking occurs. Based on the normalization condition, $\Omega_{r_c}$ is also related to $\Omega_m$ and $\Omega_k$ as $(\sqrt{\Omega_m+\Omega_{r_c}}+\sqrt{\Omega_{r_c}})^2+\Omega_k=1$.

Summarizing, the Friedmann equations of the all the models presented in the this section will be used to calculate the angular diameter distance
\begin{eqnarray}
D_A(z;\mathbf{p})=\frac{c}{1+z}\int^z_0\frac{1}{H(z')}dz',
\end{eqnarray}
where $\mathbf{p}$ denotes relevant cosmological model parameters, i.e., $\mathbf{p}\,=\,[H_0,\Omega_m]$ for flat cosmological models and
$\mathbf{p}\,=\,[H_0,\Omega_m,\Omega_k]$ for non-flat cosmological models.

\section{Observational quasar data and Methodology} \label{sec:data}

Here we work with large-scale data, selecting and combining complete samples of quasars to investigate the late-time Universe. In this section, we describe each data set and the methodology used for the cosmological analyses. These are carried out using the probes separately, followed by the joint analysis.

\textit{Distance measurements from lensed quasars.---}
In strong lensing systems with quasars acting as background sources, the time difference (time delay) between two images of the source depends on the time-delay distance $D_{\mathrm{\Delta t}}$ and the gravitational potential of the lensing galaxy by \citep{1990CQGra1319P,1990CQGra1849P,2016A&ARv..24...11T}
\begin{equation}
\Delta t_{i,j} = \frac{D_{\mathrm{\Delta t}}}{c}\Delta \phi_{i,j}(\mathbf{\xi}_{lens}).
\label{relation}
\end{equation}
Here $c$ is the speed of light, and $\Delta\phi_{i,j}=[(\mathbf{\theta}_i-\mathbf{\beta})^2/2-\psi(\mathbf{\theta}_i)-(\mathbf{\theta}_j-
\mathbf{\beta})^2/2+\psi(\mathbf{\theta}_j)]$ represents the Fermat potential difference between the image $i$ and image $j$, which is determined by the lens model parameters $\mathbf{\xi}_{lens}$ inferred from high resolution imaging of the host arcs. $\mathbf{\theta}_{i}$ and $\mathbf{\theta}_{j}$ are the angular positions of the image $i$ and $j$ in the lens plane. It is worth noting that the line-of-sight (LOS) mass distribution to the lens could also affect the Fermat potential inference, the contribution of which requires deep and wide field imaging of the area around the lens system. The two-dimensional lensing potential at the image positions, $\psi(\mathbf{\theta}_i)$ and $\psi(\mathbf{\theta}_j)$, and the unlensed source position $\mathbf{\beta}$ can be determined by the lens mass model. The time-delay distance $D_{\mathrm{\Delta t}}$ is a combination of three angular diameter distances expressed as
\begin{equation}
D_{\mathrm{\Delta
t}}\equiv(1+z_d)\frac{D^A_{\mathrm{d}}(z;\mathbf{p})
D^A_{\mathrm{s}}(z;\mathbf{p})}{D^A_{\mathrm{ds}}(z;\mathbf{p})},
\end{equation}
where the superscript ($A$) denotes the angular diameter distance, while the subscripts ($d$ and $s$) represent the deflector (or lens) and the source, respectively.

Moreover, the angular diameter distance to the deflector (or lens) can be independently inferred from the kinematic modeling with additional information of the lensing galaxy. The measured velocity dispersion provides the depth of gravitational potential at the lensing position, while the time delay provides the mass of the lensing galaxy enclosed within the position at which images are formed. Therefore, the combination of the above two quantities will generate the physical size of the system, on the base of which one could obtain the measurement of $D_d$ at the lens position divided by the angular separation of lensed images. More specifically, by choosing a suitable mass density profile (such as the power-law lens distribution) and combine it with the kinematic information of the lensing galaxy (the light distribution function $\mathbf{\xi}_{light}$, the projected stellar velocity dispersion $\sigma_P$, and the anisotropy distribution of the stellar orbits $\beta_{ani}$), one can obtain the angular diameter distance to the lens
\begin{equation}
D_{d}=\frac{1}{1+z_d}D_{\Delta t}\frac{c^2H(\mathbf{\xi}_{lens},\mathbf{\xi}_{light},\beta_{ani})}{\sigma^2_P},
\end{equation}
where the function $H$ captures all of the model components calculated from the sky angle (from the imaging data) and the anisotropy distribution of the stellar orbit (from the spectroscopy). Here, we summarise the crucial points required by the present work and one could refer to \citet{Birrer16,Birrer19} for more details. Note that the cosmological constraints obtained from the $D_d$ sample are generally weaker than those from the $D_{\Delta t}$ sample. However, the previous analysis has also demonstrated its potential in breaking the possible degeneracies among cosmological parameters, particularly those between cosmic curvature and the redshift-varying equation of state of dark energy in some non-flat dark energy models \citep{Jee16}.

The latest sample of strong-lensing systems with time-delay observations, recently released by the H0LiCOW collaboration consist of six lensed quasars covering the redshift range of $0.654<z_s<1.789$ \citep{Wong:2019kwg}: B1608+656 \citep{Suyu10,Jee19}, RXJ1131-1231 \citep{Suyu13,Suyu14,Chen19}, HE0435-1223 \citep{Wong17,Chen19}, SDSS1206+4332 \citep{Birrer19}, WFI2033-4723 \citep{Rusu20}, and PG1115+080 \citep{Chen19}. The redshifts of both lens and source, the time-delay distances, and the angular diameter distance to the lenses for these lensed quasar systems are summarized in Table 2 of \citet{Wong:2019kwg}. Note that for the lens of B1608+656, its $D_{\Delta t}$ measurement is given in the form of skewed log-normal distribution (due to the absence of blind analysis of relevant cosmological quantities), while the derived $D_{\Delta t}$ for other five lenses are given in the form of Monte Carlo Markov chains (MCMC). As for the measurements of the angular diameter distance to the the lens $D_d$, only four strong lensing systems (B1608+656, RXJ1131-1231, SDSS1206+4332, and PG1115+080) are used in our statistical analysis, which are provided in the form of MCMC. We remark here that a kernel density estimator is used to compute the posterior distributions of $\mathcal{L}_{(D_{\Delta t}, D_{d})}$ or $\mathcal{L}_{D_{d}}$ from chains, which allows to account for any correlations between $D_{\Delta t}$ and $D_{d}$ in $\mathcal{L}_{(D_{\Delta t}, D_{d})}$. The distances posterior distributions for six time delay distances (denoted as "$6D_{\Delta t}$" for simplicity) and four angular diameter distances to the lenses (denoted as "$4D_{d}$" for simplicity) are available on the website of H0LiCOW \footnote{http://www.h0licow.org}. For more works in cosmology by using strong lensing time delays, we refer the reader to see the literature  \citep{2021MNRAS.503.1096D,2022ApJ...927..191B,2021A&A...656A.153S,2021ApJ...906...26L,
2015A&A...580A..38R}.

\textit{Distance measurements from radio quasars.---} Being the brightest sources in the universe, quasars have considerable potential to be used as useful cosmological probes \citep{Liu20AJ,LiuT21a}, despite of the extreme variability in their luminosity and high observed dispersion. For instance, \citet{Risaliti18} attempted to use quasars as standard candles through the non-linear relation between their intrinsic UV emission from an accretion disk and the X-ray emission from hot corona, through the analysis and refinement of the quasar sample with well-measured X-ray and UV fluxes. In this work, we focus on the "angular size-distance" relation of ultra-compact structure in radio quasars that can be observed up to high redshifts, with milliarcsecond angular sizes measured by the very-long-baseline interferometry (VLBI). Specially, with the the signal received at multiple radio telescopes across Earth's surface, together with the registered correlated intensities considering the different arrival times at various facilities, the characteristic angular size of a distant radio quasar is defined as
\begin{equation}
\theta={2\sqrt{-\ln\Gamma \ln 2} \over \pi B} \label{thetaG}
\end{equation}
where $B$ is the interferometer baseline measured in wavelengths and the visibility modulus $\Gamma=S_c/S_t$ is the ratio between the
total and correlated flux densities. With gradually refined selection technique and the elimination of possible systematic errors, \citet{Cao:2017abj} compiled a sample of 120 intermediate-luminosity radio quasars ($10^{27}W/Hz < L < 10^{28} W/Hz$) with reliable measurements of the angular size of the compact structure from updated VLBI observations. It is now understood that the dispersion in linear size is greatly mitigated \citep{Cao:2016dgw}, i.e., the linear sizes of these standard rulers show negligible dependence on both redshifts and intrinsic luminosity. Our quasar data come from a newly compiled sample of these standard rulers from observations of 120 intermediate-luminosity quasars with angular sizes $\theta(z)$ and redshifts $z$ listed in Table 1 of \citet{Cao:2017ivt}, which extend the Hubble diagram to a redshift range $0.46<z<2.76$ currently inaccessible to the traditional resorts.

The corresponding theoretical predictions for the angular sizes of the compact structure can be written as
\begin{equation}
\theta(z)=\frac{l_m}{D^A(z;\mathbf{p})},
\end{equation}
where $D^A(z)$ is the angular diameter distance at redshift $z$, which is related to different combinations of cosmological parameters $\mathbf{p}$ (Hubble constant and the dimensionless expansion rate expansion rate). The intrinsic length $l_m$ needs to be calibrated with external indicators such SNe Ia. In this analysis, we adopt the calibration results of such quantity $l_m=11.03\pm0.25$ pc through a new cosmology-independent technique, by using the Gaussian Process to reconstruct the expansion
history of the Universe from 24 cosmic chronometer measurements \citep{Cao:2017abj}. The data of ultra-compact structures in radio quasars, which is denoted as "QSO" in this work, has been extensively used for cosmological applications in the literature \citep{Li17,Cao:2017abj,Qi17,Ma17,Cao:2016dgw}.

Now the posterior likelihood $\mathcal{L}_{QSO}$ for can be constructed by
\begin{eqnarray}
\mathcal{L}_{QSO}=\prod_{i=1}^{120}\frac{1}{\sqrt{2\pi(\sigma_{i}^2+\sigma_{sys}^2)}}
\times\exp\bigg\{
-\frac{\left[\theta(z_{i};\mathbf{p})
     - \theta_{obs,i}\right]^{2}}{2(\sigma_{sta}^{2}+\sigma_{sys}^2)}\bigg\},
\end{eqnarray}
where $\theta_{obs,i}$ is the observed angular size for the \textit{i}th data point in the sample, $\sigma_{sta}$ is the observational statistical uncertainty for the \textit{i}th quasar. According
to the error strategy proposed in \citet{Cao:2017ivt}, an additional $10\%$ systematic uncertainty ($\sigma_{sys}=0.1\theta_{obs}$) in the observed angular sizes is also considered, accounting
for the intrinsic spread in the linear size \citep{Cao:2016dgw,Qi2020arx}. Such strategy has been extensively applied in the subsequent cosmological studies with such standard ruler data, which extended our understanding of the evolution of the Universe to $z\sim3$ \citep{Ryan21,Vavry20,Melia18}.

Summarizing, we use different combinations of lensed and unlensed quasars by adding the log-likelihood of cosmological parameters $H_0$, $\Omega_m$ and $\Omega_k$ (if available) for the posterior distributions of six lens time-delay distances $\mathcal{L}_{D_{\Delta t}}$, four angular diameter distances to the lenses $\mathcal{L}_{D_{d}}$, and 120 angular diameter distances to the radio quasars $\mathcal{L}_{QSO}$.
The final log-likelihood
\begin{equation}
\mathrm{ln} {\cal L}= \mathrm{ln} ({\cal L}_{\rm{QSO}})+\mathrm{ln}
({\cal L}_{(D_{\Delta t},D_{d})}),
\end{equation}
is sampled in the framework of Python MCMC module EMCEE \citep{Foreman_Mackey_2013}. Meanwhile, in this analysis we introduce the DIC to evaluate which model is more consistent with the observational data, focusing on the Deviance Information Criterion (DIC) to compare the goodness of fit on models with different numbers of parameters \citep{Spiegelhalter03}. The DIC is defined as
\begin{equation}
DIC=D(\bar{\theta})+2p_D=\overline{D(\theta)}+p_D,
\end{equation}
where $D(\theta)=-2\mathrm{ln}{\cal L(\theta)}+C$, $C$ is a normalized constant depending only on the data that disappears from ant derived quantity, and $p_D=\overline{D(\theta)}-D(\bar{\theta})$ is the effective number of model parameters, with the deviance of the likelihood $D$. Specially, when we use $\chi^2=-2\mathrm{ln}{\cal L(\theta)}$ to describe the $p_D$, it can be rewritten as
\begin{equation}
p_D=\overline{\chi^2(\theta)}-\chi^2(\bar{\theta}).
\end{equation}
Compared with the widely-used Akaike Information Criterion (AIC) or Bayesian Information Criterion (BIC), the advantage of DIC lies in the fact that it is determined by the qualities which can be easily obtained from Monte Carlo posterior samples. Moreover, parameters that are unconstrained by the data would also be appropriately treated in the framework of DIC \citep{Liddle07}.

\begin{table*}
\renewcommand\arraystretch{1.5}
\caption{The best-fit values and 1$\sigma$ uncertainties for the parameters ($H_0$, $\Omega_m$, $\Omega_{r_c}$ (if available), and $\Omega_k$) in each cosmological model and quasar data set. The $\chi^2$ and DIC values for all models are also added for comparison.}
\begin{center}
\begin{tabular}{l| c c c c c c c}
\hline
\hline
Model  & Data set & $H_0 (\Mpc)$ &$\Omega_m$ & $\Omega_k$  & $\Omega_{r_c}$  & $\chi^2$ & DIC  
\\
\hline
Flat $\Lambda$CDM  &  6$D_{\Delta t}$  & $ 73.26^{+1.74}_{-1.84}$ & $0.297^{+0.133}_{-0.127}$ & $ -$& $ -$& 39.98  &42.64
\\

&  6$D_{\Delta t}$+QSO  & $73.51^{+1.73}_{-1.82}$ & $0.281^{+ 0.050}_{-0.041}$ & $ -$& $ -$&358.99 &364.21 
\\

& 6$D_{\Delta t}$+$4D_d$+QSO  & $73.42^{+1.74}_{-1.84}$ & $0.285^{+0.049}_{-0.041}$ & $ -$& $ -$& 385.38 &$ 391.33$ 
\\

\hline

Flat PEDE    &  6$D_{\Delta t}$  & $74.94^{+1.89}_{-2.03}$ & $0.259^{+0.173}_{-0.137}$ & $-$& $ -$ & 39.57 &42.76
\\

&  6$D_{\Delta t}$+QSO  & $ 75.13^{+1.87}_{-2.07}$ & $0.294^{+0.046}_{-0.040}$ & $-$& $ -$& 358.34 &363.81 
\\

&  6$D_{\Delta t}$+$4D_d$+QSO  & $75.08^{+1.83}_{-2.05}$ & $0.297^{+0.046}_{-0.038}$ & $ -$& $-$&384.53 &390.68  
\\

\hline
Flat DGP  &  6$D_{\Delta t}$  & $67.75^{+2.43}_{-3.95}$ & $0.251^{+ 0.184}_{-0.137}$ & $ -$& $ 0.140^{+0.056}_{-0.059}$&40.43 &45.74
\\

&  6$D_{\Delta t}$+QSO  & $67.84^{+ 1.12}_{-1.19}$ & $ 0.243^{+ 0.043}_{-0.037}$ & $-$& $ 0.143^{+0.015}_{-0.016}$&360.02 &367.72 
\\

&  6$D_{\Delta t}$+$4D_d$+QSO  & $ 67.84^{+1.10}_{-1.12}$ & $0.246^{+ 0.043}_{-0.037}$ & $ -$& $ 0.142^{+0.014}_{-0.016}$&386.42  &393.67
\\

\hline
Non-flat $\Lambda$CDM  &  6$D_{\Delta t}$  & $74.38^{+2.12}_{-2.34}$ & $0.242^{+0.164}_{-0.129}$ & $ 0.258^{+ 0.167}_{-0.253}$& $ -$&39.80  &44.67
\\

&  6$D_{\Delta t}$+QSO  & $73.39^{+ 2.06}_{-2.11}$ & $  0.274^{+0.083}_{-0.080}$ & $ 0.036^{+ 0.223}_{-0.214}$& $ -$&358.31 &366.89
\\

& 6$D_{\Delta t}$+$4D_d$+QSO  & $73.78^{+1.99}_{-2.17}$ & $0.254^{+0.083}_{-0.074}$ & $0.100^{+0.214}_{-0.213}$ & $ -$& 384.69 & 393.81
\\

\hline

Non-flat PEDE    &  6$D_{\Delta t}$  &  $74.84^{+15.81}_{-8.72}$ & $ 0.247^{+0.180}_{-0.131}$ & $ -0.027^{+ 0.340}_{-0.306}$& $ -$&39.49  &44.23
\\

&  6$D_{\Delta t}$+QSO  & $71.67^{+ 16.75}_{-7.38}$ & $  0.315^{+0.094}_{-0.106}$ & $ -0.089^{+ 0.365}_{-0.276}$& $ -$ &357.82 &366.39
\\

&  6$D_{\Delta t}$+$4D_d$+QSO  &  $75.05^{+2.01}_{-2.19}$ & $0.275^{+0.081}_{-0.077}$ & $0.076^{+0.222}_{-0.201}$ & $ -$& 383.81 & 393.45 
\\

\hline
Non-flat DGP  &  6$D_{\Delta t}$  & $68.31^{+ 15.37}_{-10.45}$ & $0.251^{+ 0.186}_{-0.142}$ & $ 0.023^{+ 0.316}_{-0.345}$& $ 0.135^{+0.143}_{-0.116}$&40.31  &46.24
\\

&  6$D_{\Delta t}$+QSO  &  $66.26^{+ 13.57}_{-8.04}$ & $   0.253^{+0.089}_{-0.081}$ & $ -0.050^{+ 0.328}_{-0.305}$& $ 0.151^{+0.107}_{-0.101}$ &359.47 & 367.72
\\

&  6$D_{\Delta t}$+$4D_d$+QSO  &  $ 66.35^{+14.92}_{-8.10}$ & $0.253^{+0.086}_{-0.083}$ & $-0.047^{+0.348}_{-0.311}$& $ 0.150^{+0.110}_{-0.104}$&385.19  &395.27
\\
\hline
\hline

\end{tabular}
\end{center}
\end{table*}\label{tab:res_models}

\section{Results and discussion} \label{sec:res}


\begin{figure*}
\begin{center}
\includegraphics[width=0.45\linewidth]{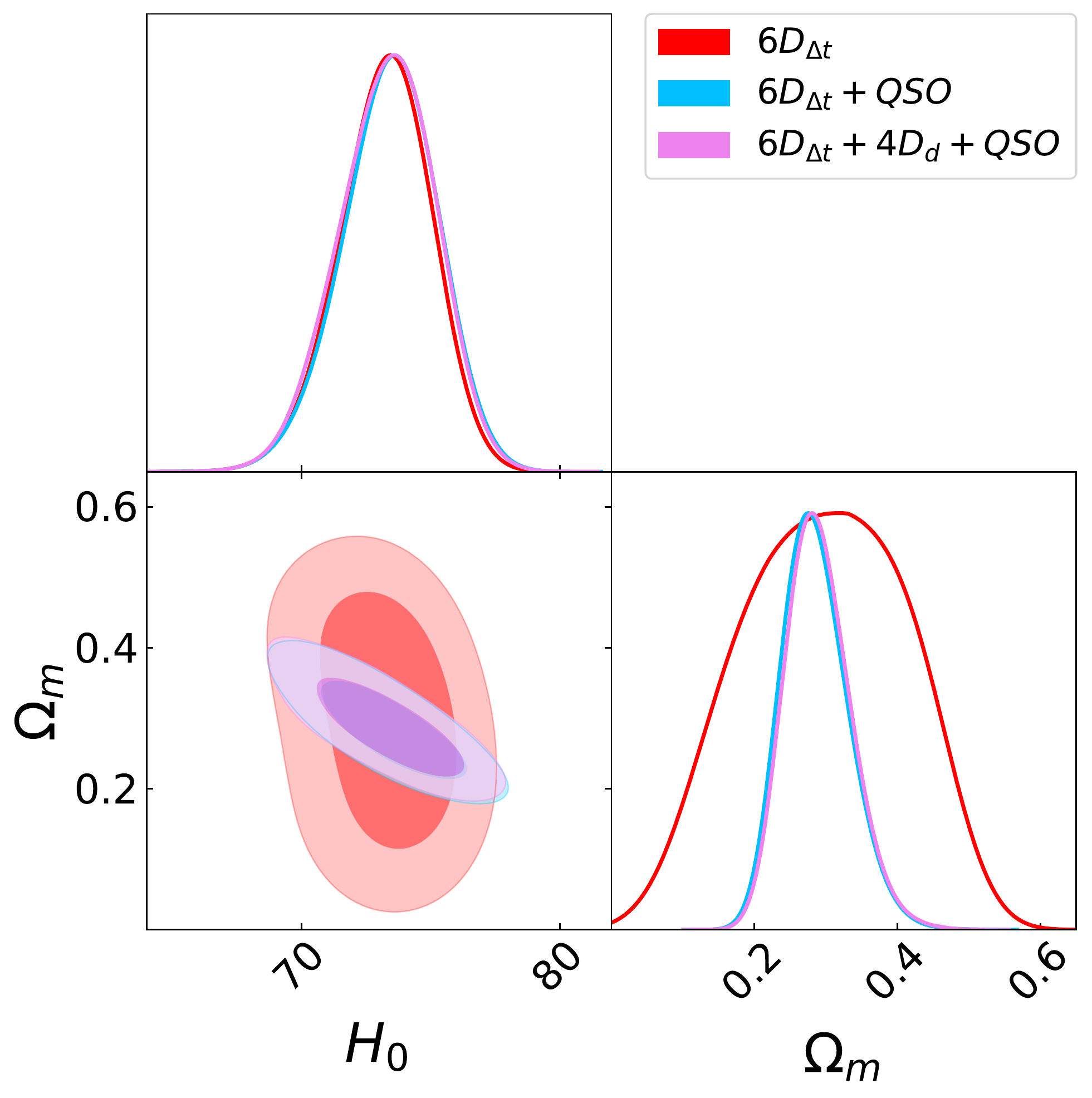} \\
\includegraphics[width=0.45\linewidth]{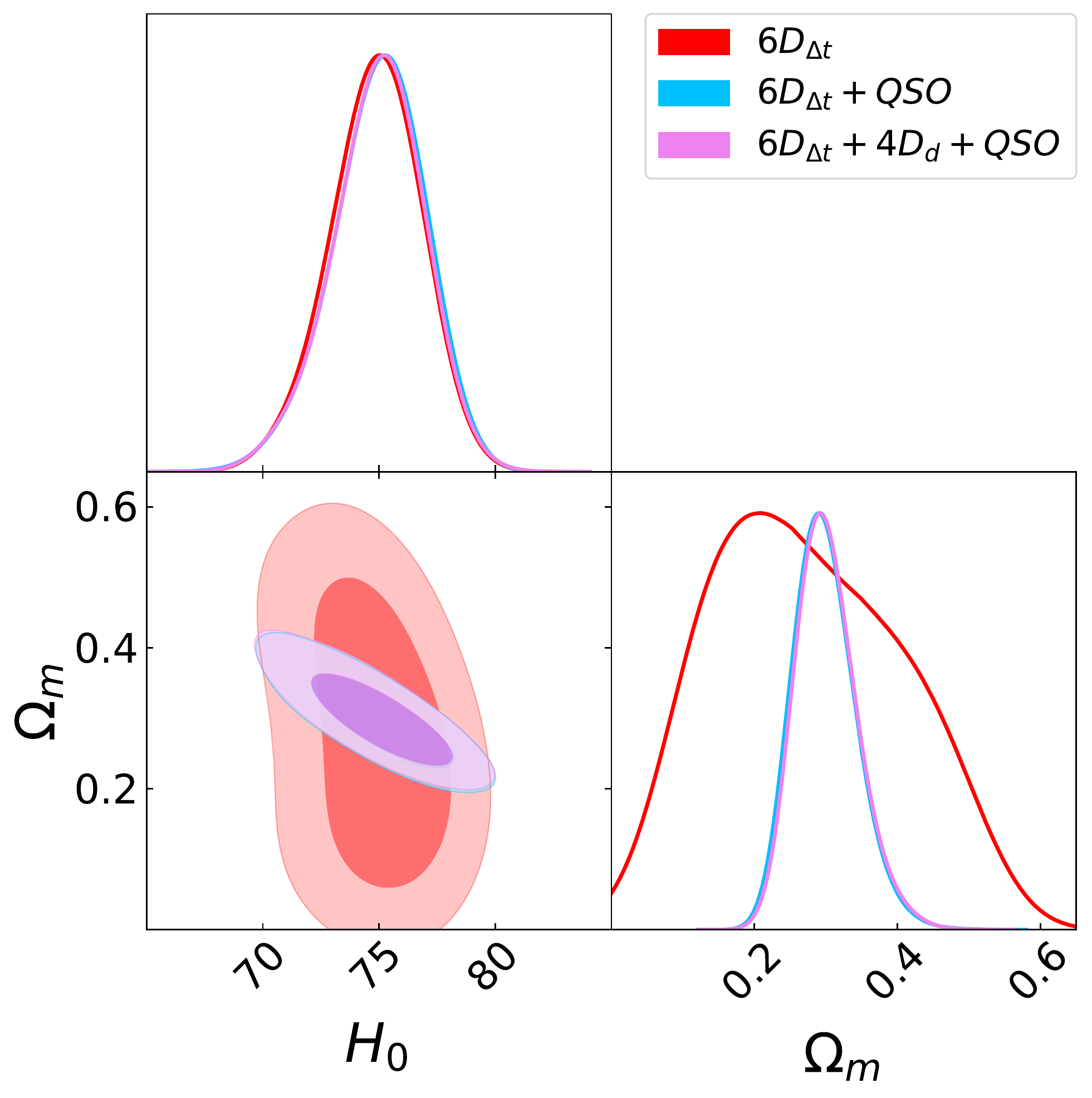}
\includegraphics[width=0.45\linewidth]{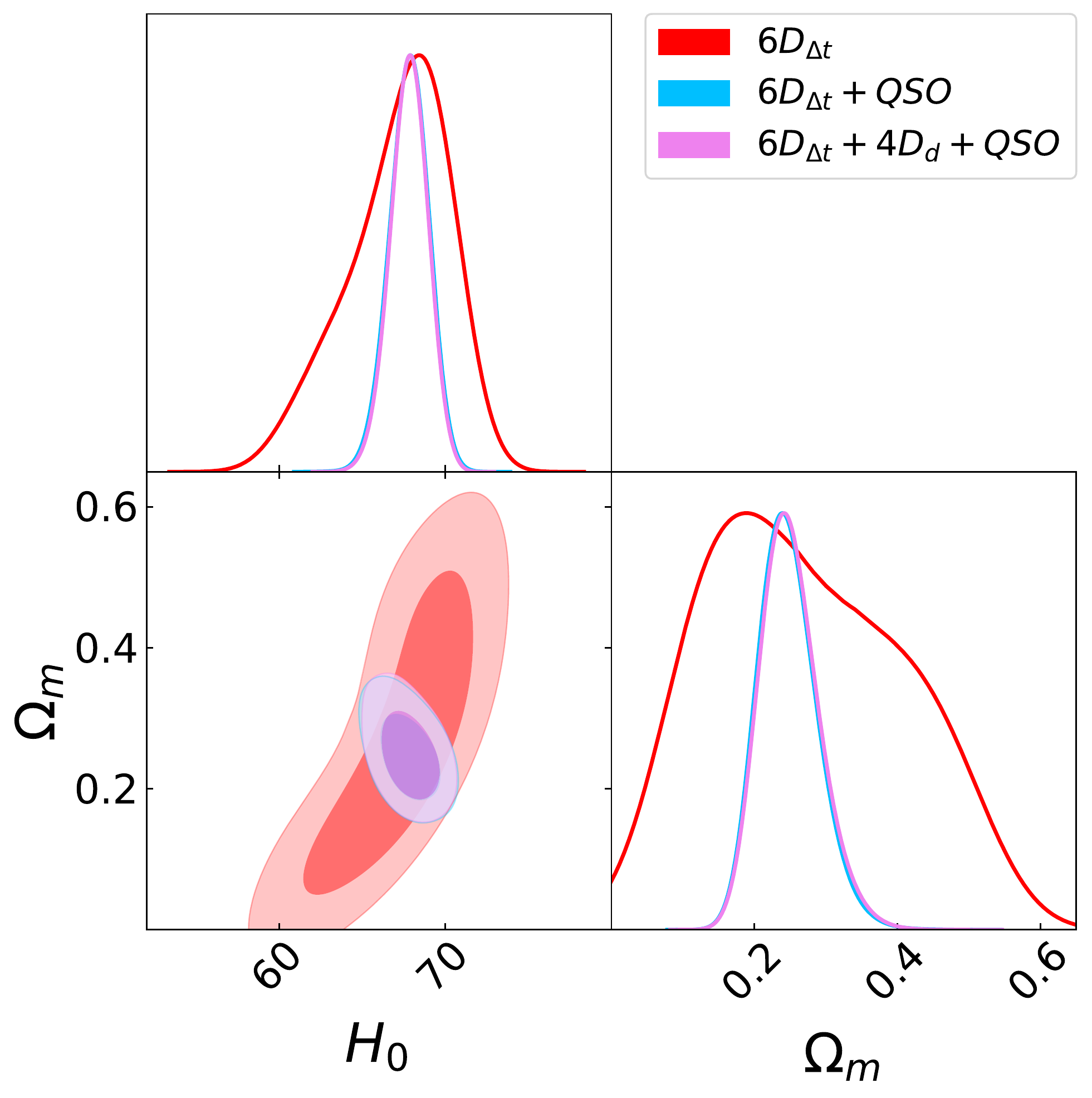}
\end{center}
\caption{The 2-D plots and 1-D marginalized distributions with 1-$\sigma$ and 2-$\sigma$ contours of cosmological parameters ($H_0$ and $\Omega_m$) in the framework of flat $\Lambda$CDM (upper), flat PEDE (lower left) and flat DGP (lower right) models with lensed quasars ($6D_{\Delta t}$, $4 D_d$) and radio quasars (QSO).
}
\label{fig:flat_contours}
\end{figure*}


\begin{figure*}
\begin{center}
\includegraphics[width=0.49\linewidth]{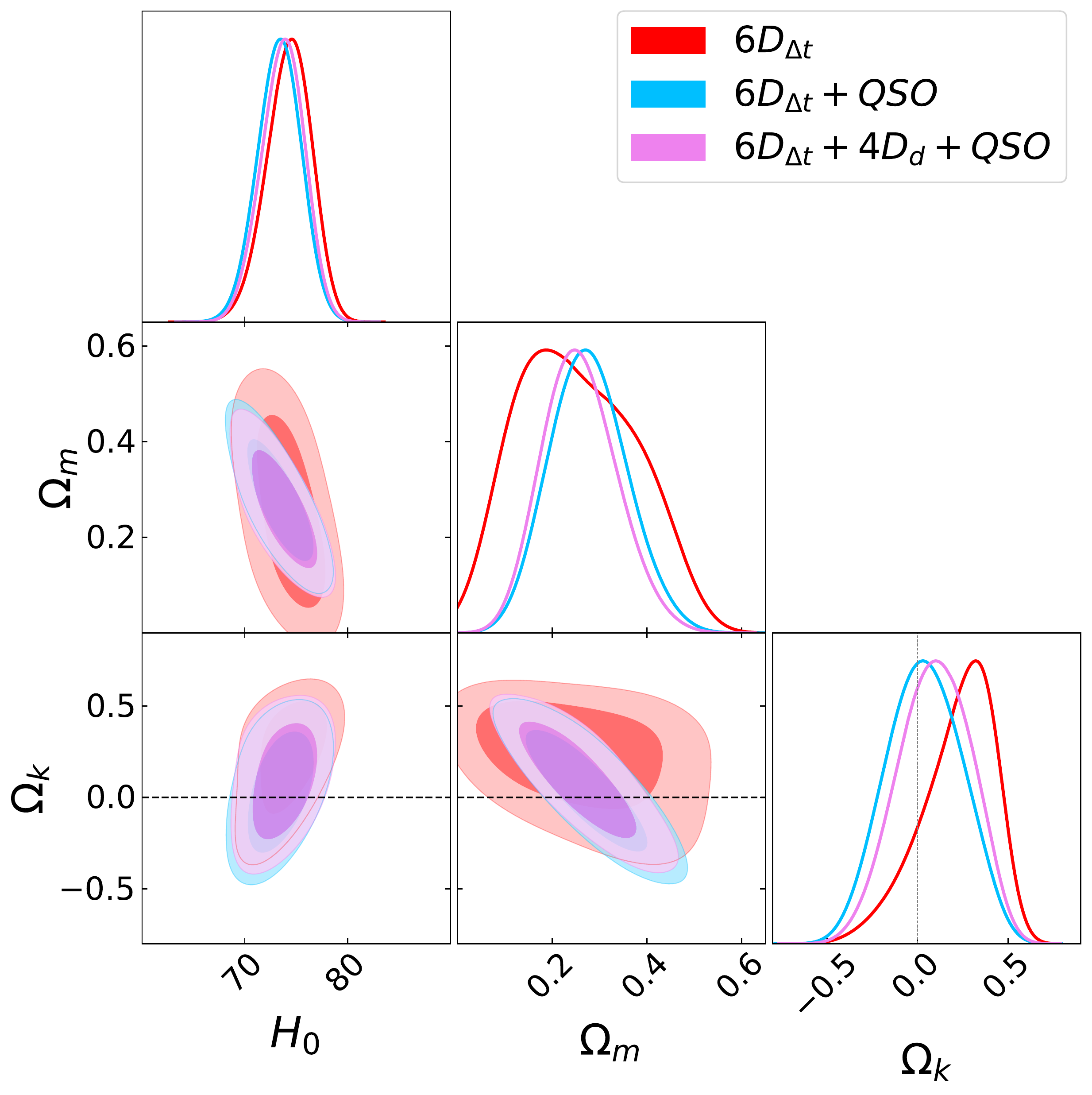}
\\
\includegraphics[width=0.49\linewidth]{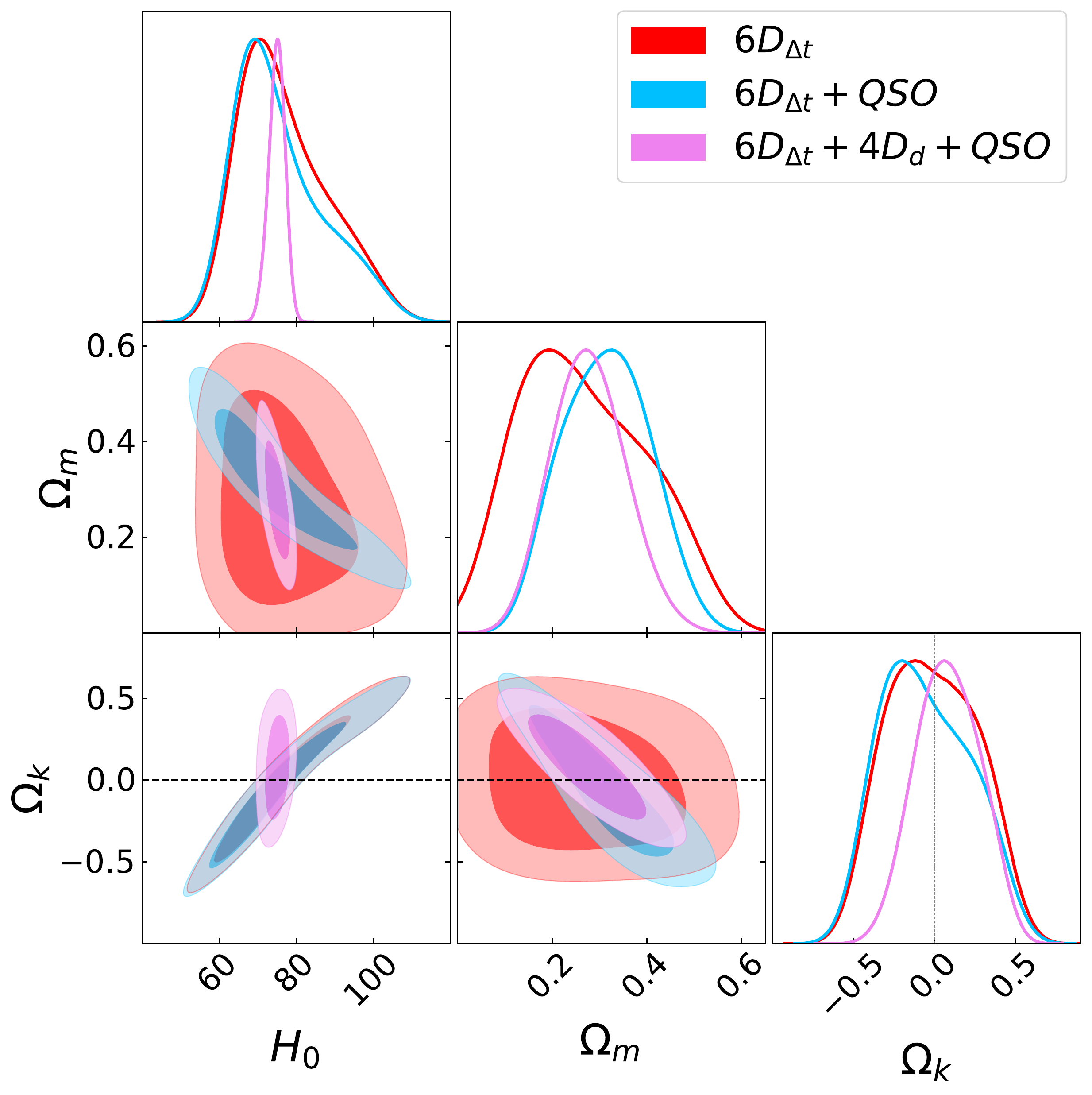}
\includegraphics[width=0.49\linewidth]{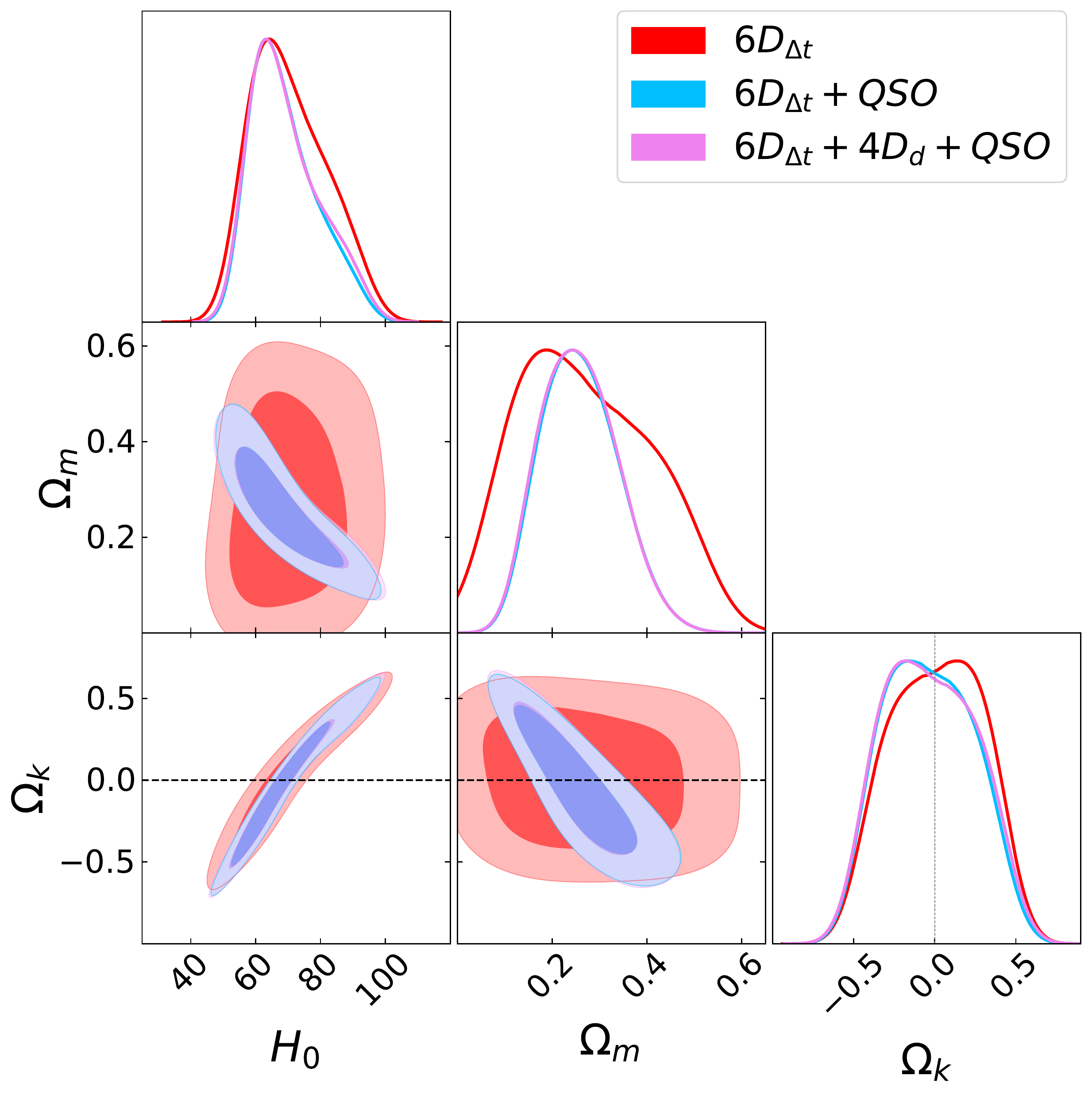}
\end{center}
\caption{The 2-D plots and 1-D marginalized distributions with 1-$\sigma$ and 2-$\sigma$ contours of cosmological parameters ($H_0$, $\Omega_m$ and $\Omega_k$) in the framework of non-flat $\Lambda$CDM (upper), non-flat PEDE (lower left) and non-flat DGP (lower right) models with lensed quasars ($6D_{\Delta t}$, $4 D_d$) and unlensed quasars (QSO). }
\label{fig:nonflat}
\end{figure*}

In order to demonstrate the constraining power of the latest observations of quasars, we use different data combinations (6$D_{\Delta t}$, 6$D_{\Delta t}$+QSO, and 6$D_{\Delta t}$+4$D_{d}$+QSO) to place constraints on cosmological parameters in $\Lambda$CDM, PEDE and DGP models both in flat and non-flat cases. The posterior one-dimensional (1-D) probability distributions and two-dimensional (2-D) confidence regions of the cosmological parameters for the six flat and non-flat models are shown in Figs.~1-2. We list the marginalized best-fitting parameters and $1\sigma$ uncertainties for all models and data combinations in Table 1. The corresponding $\chi^2$ and DIC values are also listed in Table 1.

\textit{Flat cases.---} The constraint results for three flat cosmological models are presented in Fig.~\ref{fig:flat_contours}, in which we show the 2-D confidence regions (with 1$\sigma$ and 2$\sigma$ limits) as well as 1-D marginalized distributions from different data combinations. Our findings show that the constraints we obtain from the combination of the latest observations of quasars are more reliable than the those derived from independent quasar sample. Fortunately, the results in Fig.~2 show that the radio quasar data place remarkable constraints on all parameters of the three cosmological models, and the degeneracy among different model parameters are well broken. When combined with 120 unlensed quasar data, the 6 time-delay lensed quasars would produce tighter constraints on the matter density parameter in all of the three cosmological scenarios. The measured $\Omega_m$ range from a low value of $ 0.243^{+ 0.043}_{-0.037}$ (flat DGP) to a high value of $0.294^{+0.046}_{-0.040}$ (flat PEDE). In particular, for flat $\Lambda$CDM the derived matter density parameter from 120 unlensed quasars and 6 time-delay lensed quasars $\Omega_m=0.297^{+0.133}_{-0.127}$ shows a perfect agreement with the TT, TE, EE+lowE+lensing results of Planck Collaboration ($\Omega_{m}=0.3103 \pm0.0057$) \citep{Aghanim:2018eyx}. Such findings are different from the those obtained from the latest compilation of X-ray+UV quasars acting as standard candles, which favors a larger value of the matter density parameter at higher redshifts \citep{Risaliti18}. See also \citet{Lian21} for more discussion about this issue. In spite of the low mean values of $\Omega_m$ in the flat DGP model, the constraints on $\Omega_m$ obtained with 6$D_{\Delta t}$, 6$D_{\Delta t}$+QSO, and 6$D_{\Delta t}$+QSO+4$D_{d}$ data are mutually consistent with those obtained from other astrophysical probes \citep{Xu10,Xu14,Giannantonio08}. For comparison, the corresponding fits on the parameter of $\Omega_{r_c}$ in flat DGP model are also displayed in Table 1. Finally, our analysis demonstrates that the matter density parameter plays an important role in the determination of the Hubble constant, which can be clearly seen from the anti-correlation between $\Omega_m$ and $H_0$ in Fig.~1.

The constraints on the Hubble constant are between $H_0=67.75^{+2.43}_{-3.95}$ $\Mpc$ (DGP) and $H_0=74.94^{+1.89}_{-2.03}$ $\Mpc$ (PEDE) with 6 time-delay lensed quasars, which shift to $H_0=67.84^{+ 1.12}_{-1.19}$ $\Mpc$ (DGP) and $H_0=75.13^{+1.87}_{-2.07}$ $\Mpc$ (PEDE) with the combined 6$D_{\Delta t}$+QSO data. Specially, for the flat $\Lambda$CDM and PEDE model, the mean values of $H_0$ obtained with 6$D_{\Delta t}$+4$D_d$+QSO data are more consistent with
the recent determinations of $H_0$ from the Supernovae H0 for the SH0ES collaboration \citep{Riess:2019cxk}. However, in the framework of flat DGP model, the measured value of $H_0$ with 1$\sigma$ uncertainty ($67.84^{+ 1.12}_{-1.19}$ $\Mpc$), which is 3.6$\sigma$ lower than the statistical estimates of the SH0ES results, demonstrates a perfect agreement with that derived from the recent Plank CMB observations \citep{Aghanim:2018eyx}. In addition, relative to the 6$D_{\Delta t}$+QSO constraints, the Hubble constant derived from the combined 6$D_{\Delta t}$+4$D_L$+QSO data are a little higher than those values measured from the 6$D_{\Delta t}$+QSO case. However, these differences are not statistically significant given the error bars, as can be seen from the numerical results summarized in Table.~\ref{tab:res_models}.

\textit{Non-flat cases.---} As was revealed in the recent studies of
\citet{DiValentino:2019qzk,DiValentino:2020hov,2019arXiv190809139H}, the discrepancy between the Hubble constant and cosmic curvature measured locally and inferred from Planck highlights the importance of considering non-flat cosmological models in this work. The constraint results for three flat cosmological models are presented in Fig.~\ref{fig:nonflat}, in which we show the 2-D confidence regions(with 1$\sigma$ and 2$\sigma$ limits) as well as 1-D marginalized distributions from different data combinations. The numerical results are also summarized in Table~1. One can see from the upper panel of Fig.~\ref{fig:nonflat} that, in the concordance $\LCDM$ cosmology a stringent constraint on the Hubble constant could be obtained from 6 time-delay quasar data $D_{\Delta t}$ ($H_0=74.38^{+2.12}_{-2.34}$ $\Mpc$). However, the MCMC chains failed to converge for the other two model parameters, i.e., matter density and cosmic curvature parameters ($\Omega_m$, $\Omega_k$). Such issue could be appropriately addressed with stringent constraints produced by the combination of 120 QSO sample and 6 time-delay lensed quasars, with the best-fitting values with 68.3\% confidence level for the three parameters: $H_0=73.39^{+ 2.06}_{-2.11}$ $\Mpc$, $\Omega_m=0.274^{+0.083}_{-0.080}$, and $\Omega_k=0.036^{+ 0.223}_{-0.214}$. With the combined data sets 6$D_{\Delta t}$+QSO+4$D_{d}$, we also get stringent constraints on the model parameters
$H_0=73.78^{+1.99}_{-2.17}$ $\Mpc$, $\Omega_m=0.254^{+0.083}_{-0.074}$, and  $\Omega_k=0.100^{+0.214}_{-0.213}$. Compared with the previous results obtained in other model-independent methods \citep{Qi:2018aio}, our analysis results also demonstrate the strong degeneracy between the Hubble constant, the matter density parameter and cosmic curvature, which would be effectively broken by the combination of the latest observations of quasars, i.e., the angular size of compact structure in radio quasars as standard rulers and the time delays from gravitationally lensed quasars. The combination of the quasar data sets, justified by their consistency within $1\sigma$, retains the same correlation between $H_0$ and $\Omega_k$ as distinct samples of quasars. For the determination of Hubble constant, our constraint in the framework of non-flat $\Lambda$CDM cosmology is well consistent with the local Hubble constant measurement from SH0ES collaboration \citep{Riess:2019cxk}. The determination of $\Omega_k$ suggests no significant deviation from flat spatial hypersurfaces, although favoring a somewhat positive value in the non-flat $\Lambda$CDM case.

In the case of non-flat PEDE model, it can be clearly seen from the comparison plots presented in Fig.~\ref{fig:nonflat} that there is a consistency between 6$D_{\Delta t}$, 6$D_{\Delta t}$+QSO, and 6$D_{\Delta t}$+4$D_d$+QSO data sets.
Our results confirm that the combination of compact structure in radio quasars  and angular diameter distances to the lenses (4$D_d$) could break the degeneracy between cosmological parameters and lead to a more stringent constraints on all of the cosmological parameters, which is the most unambiguous result of the current lensed + unlensed quasar data set. Interestingly, the 6$D_{\Delta t}$+QSO data generate a higher matter density parameter $\Omega_m=0.315^{+0.094}_{-0.106}$ compared with other quasar samples ($\Omega_m= 0.247^{+0.180}_{-0.131}$ for 6$D_{\Delta t}$ and $\Omega_m=0.275^{+0.081}_{-0.077}$ for 6$D_{\Delta t}$+4$D_d$+QSO). For the Hubble constant inferred from 6$D_{\Delta t}$, 6$D_{\Delta t}$+QSO, and 6$D_{\Delta t}$+4$D_d$+QSO data sets, it's obvious that the combination of 6$D_{\Delta t}$ with 120 unlensed quasar data will result in a lower $H_0$, but adding 4$D_d$ data to the combination would increase the median value of $H_0$ comparing to the $H_0$ values obtained from 6$D_{\Delta t}$ alone. The estimated values of the Hubble constant in the non-flat PEDE model are between $H_0=71.67^{+ 16.75}_{-7.38}$ $\Mpc$ and $H_0=75.05^{+2.01}_{-2.19}$ $\Mpc$ with 6$D_{\Delta t}$ and 6$D_{\Delta t}$+4$D_d$+QSO data, which
in broad terms agree very well with the standard ones reported by the SH0ES collaboration \citep{Riess:2019cxk}. For the determination of cosmic curvature in the non-flat PEDE model, our results show that there is no significant evidence indicating its deviation from zero (spatially flat geometry). More specifically, the two quasar samples of 6$D_{\Delta t}$ and 6$D_{\Delta t}$+QSO favor closed geometry ($\Omega_k=-0.027^{+ 0.340}_{-0.306}$, $\Omega_k=-0.089^{+ 0.365}_{-0.276}$), while an open universe is favoured by 6$D_{\Delta t}$+4$D_d$+QSO data sets with $\Omega_k= 0.076^{+0.222}_{-0.201}$.

Finally, we perform a comparative analysis of the current lensed+unlensed
quasar data set in the non-flat DGP model. The $1\sigma$ and $2\sigma$ confidence level contours for parameter estimations are shown in Fig.~2, with
the marginalized best-fitting parameters and $1\sigma$ uncertainties summarized in Table~1. Let us note that for such modified gravity model gravity arising from the braneworld theory, the best-fitting matter density parameter will be considerably shifted to a lower value. Our final assessments of the matter density with the corresponding 1$\sigma$ uncertainties ($\Omega_m=0.251^{+ 0.186}_{-0.142}$, $\Omega_m=0.253^{+0.089}_{-0.081}$, and $\Omega_m=0.253^{+0.086}_{-0.083}$ with 6$D_{\Delta t}$, 6$D_{\Delta t}$+QSO, and 6$D_{\Delta t}$+4$D_d$+QSO data sets) are consistent with the standard ones reported by other astrophysical probes, such as growth factors combined with CMB+BAO+SNe Ia observations \citep{Xia09}, Hubble parameter combined with CMB+SNe Ia data \citep{Fan08}, statistical analysis of strong gravitational lensing systems \citep{Cao2011b,Cao2012b,Ma2019,LiuT20b}, and the observations of compact structure in intermediate-luminosity radio quasars \citep{Cao:2017ivt,LiuAPJ19,LiuT21b}. For comparison, the corresponding fits on the parameter of $\Omega_{r_c}$ are also displayed in Table 1. We remark here that the DGP model will reduce to the concordance $\Lambda$CDM model when $\Omega_{rc}=0$. Considering its non-vanishing value revealed in this analysis and previous works ($\Omega_{r_c}\sim 0.14$), our fitting result shows
the DGP model fails to recover and is only a bit worse than the $\Lambda$CDM under the current observational tests (see also \citet{Fan08,Cao2011b}). We also investigate how sensitive our results on $H_0$ and $\Omega_k$ are on the choice of this cosmological model. On the one hand, the measured value of $H_0$ with 1$\sigma$ uncertainty derived from the latest quasar sample, $H_0=68.31^{+ 15.37}_{-10.45}$ $\Mpc$ (6$D_{\Delta t}$) and $H_0=66.26^{+ 13.57}_{-8.04}$ $\Mpc$ (6$D_{\Delta t}$+QSO) shows a perfect consistency with that derived from the recent Planck CMB observation \citep{Aghanim:2018eyx}. On the other hand, our final assessments of the cosmic curvature with corresponding 1$\sigma$ uncertainty are $\Omega_k=0.023^{+0.316}_{-0.345}$ and $\Omega_k=-0.050^{+0.328}_{-0.305}$ with the two quasar samples (6$D_{\Delta t}$ and 6$D_{\Delta t}$+QSO), which are more consistent with flat spatial hypersurfaces than what \citet{Qi:2018aio} found. From the joint analyses with lensed+unlensed quasar data, we find that the Hubble constant and the spatial curvature parameter are constrained to be $H_0=66.35^{+14.92}_{-8.10}$ $\Mpc$ and $\Omega_k=-0.047^{+0.348}_{-0.311}$, which furthermore confirms the above conclusions.

\begin{figure}
\begin{center}
\includegraphics[width=0.94\linewidth]{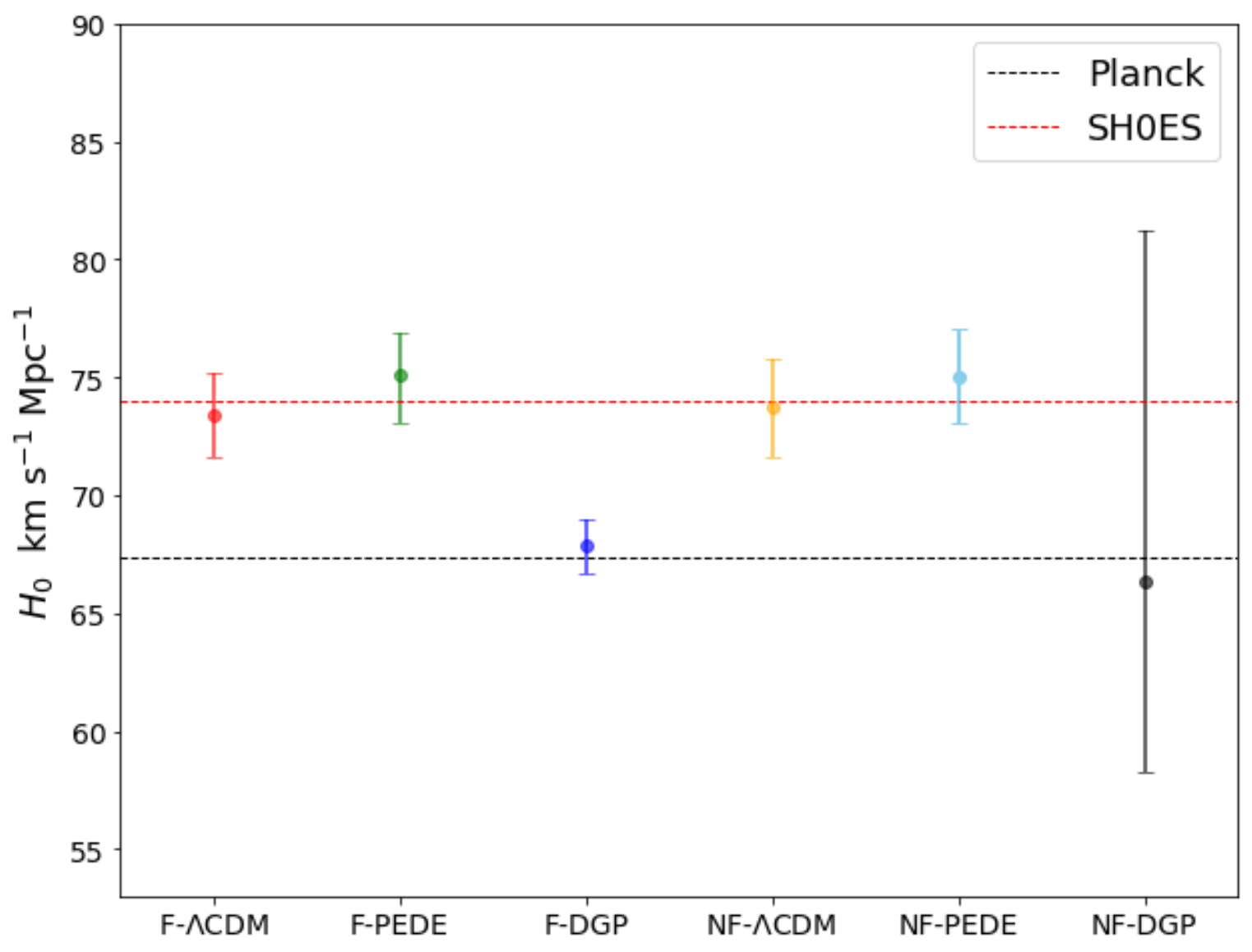}
\end{center}
\caption{Determination of Hubble constant from the combined data of lensed quasars ($6D_{\Delta t}$ and $4 D_d$) and unlensed radio quasars (QSO), in the framework of six spatially flat (F) and non-flat (NF) cosmological models ($\Lambda$CDM, PEDE and DGP). }
\label{fig:H0}
\end{figure}

\begin{figure}
\begin{center}
\includegraphics[width=0.92\linewidth]{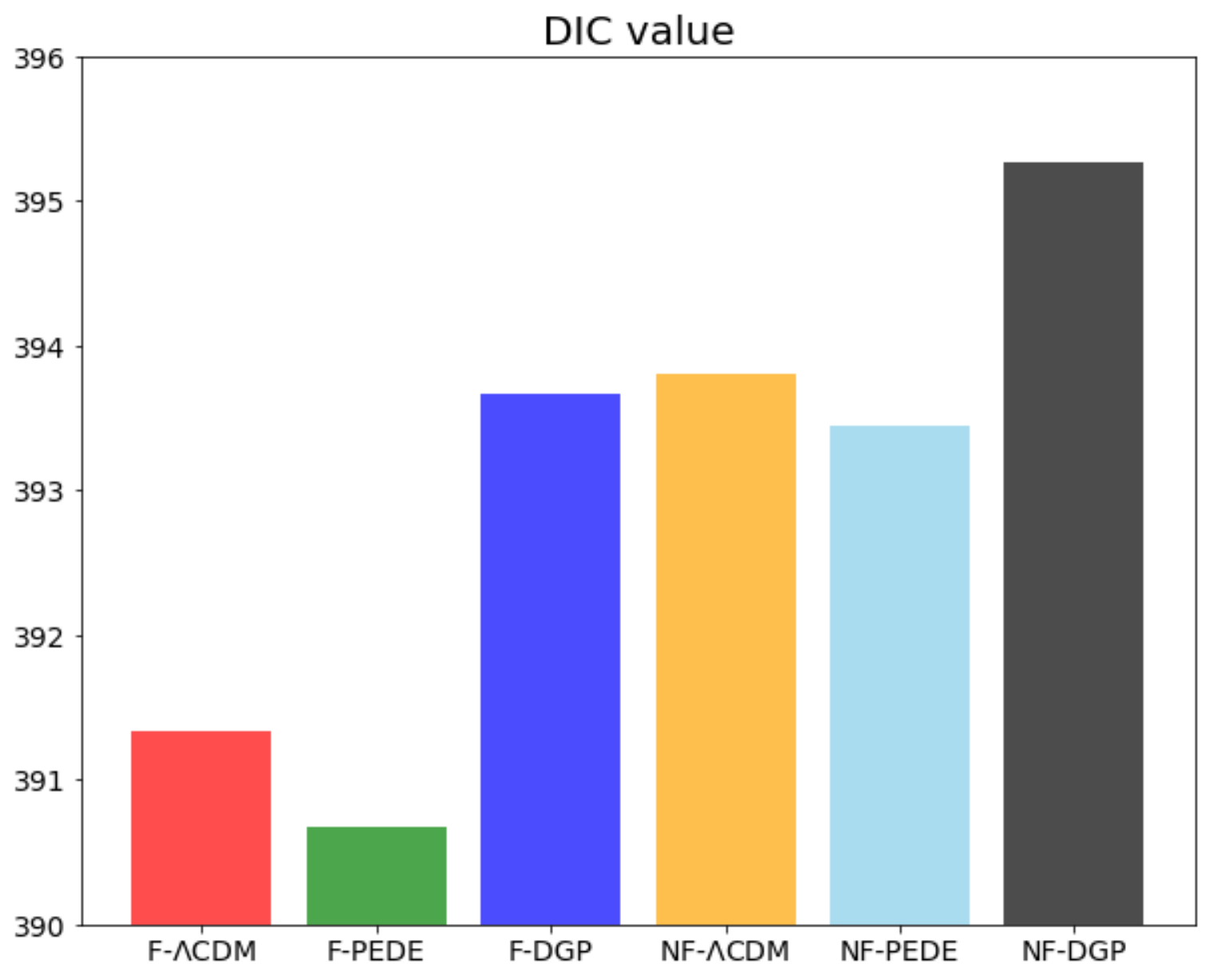}
\end{center}
\caption{Graphical representation of the DIC values for six spatially flat (F) and non-flat (NF) cosmological models ($\Lambda$CDM, PEDE and DGP), based on the combined data of lensed quasars ($6D_{\Delta t}$ and $4 D_d$) and unlensed radio quasars (QSO).}
\label{fig:DIC}
\end{figure}

\begin{figure}
\begin{center}
\includegraphics[width=0.92\linewidth]{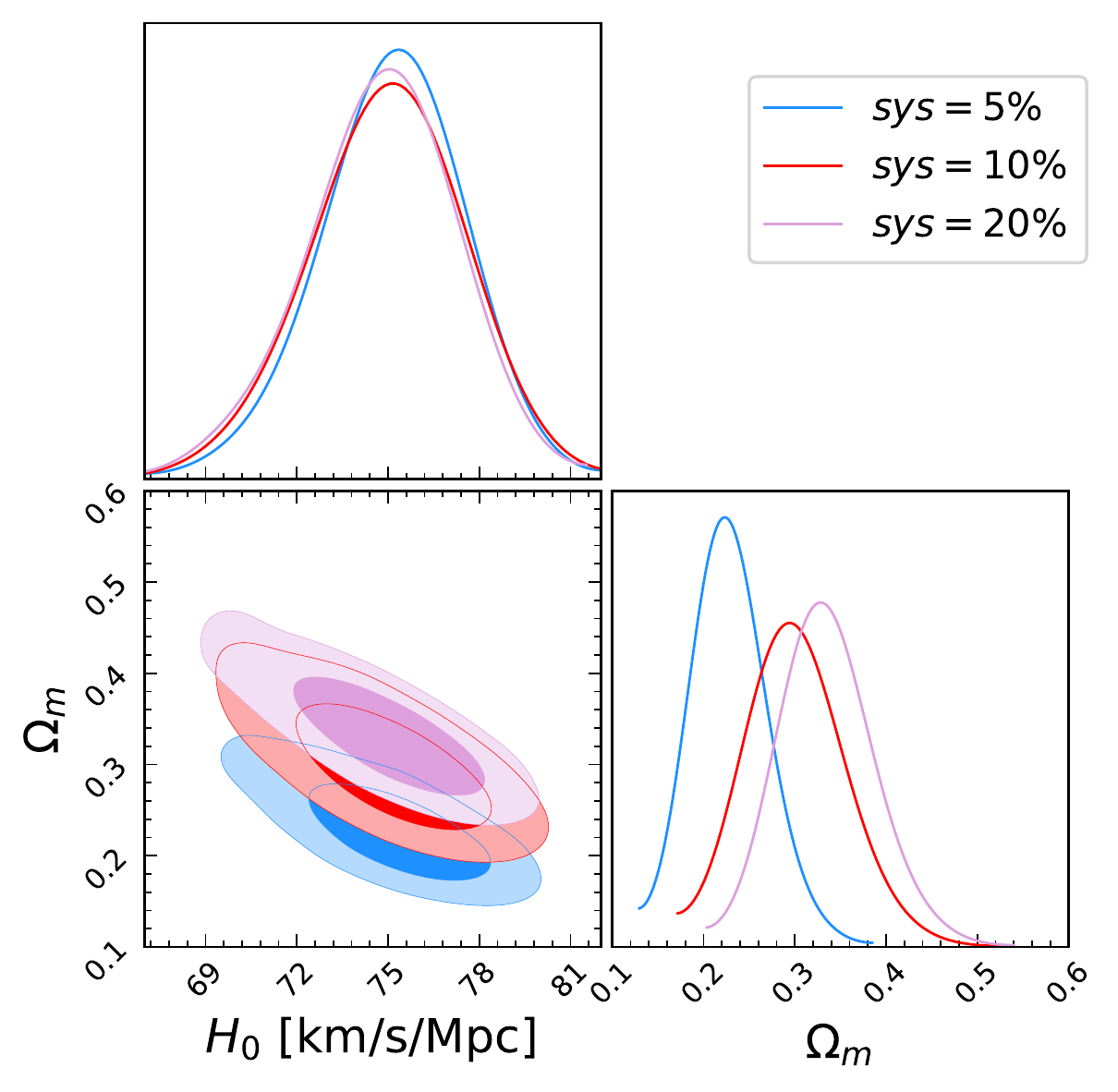}
\end{center}
\caption{Cosmological constraints on the flat PEDE model from the combined quasar data, with different systematical uncertainties in the angular size measurements of unlensed radio quasars (QSO).}
\label{fig:sys}
\end{figure}

Now let us remark on the $H_0$ and $\Omega_k$ measurements from the newly compiled sample of ultra-compact structure in radio quasars and strong gravitational lensing systems with quasars acting as background sources. The $H_0$ determination from six spatially flat and non-flat cosmological models ($\Lambda$CDM, PEDE and DGP) are displayed in Fig.~\ref{fig:H0}. A joint analysis of the quasar sample (the time-delay measurements of six strong lensing systems, four angular diameter distances to the lenses and 120 intermediate-luminosity quasars calibrated as standard rulers) provides model-independent estimations of the Hubble constant $H_0$, which is strongly consistent with that derived from the local distance ladder by SH0ES collaboration in the $\Lambda$CDM and PEDE model. However, in the framework of a DGP cosmology (especially for the flat universe), the measured Hubble constant is in good agreement with that derived from the the recent Planck 2018 results. Meanwhile, our results also demonstrate that zero spatial curvature is supported by the current lensed and unlensed quasar observations and there is no significant deviation from a flat universe. Finally, our results have provided independent evidence for the accelerated expansion of the Universe, with the existence of dominant dark energy density ($\Omega_m \sim 0.30$) in the framework of six cosmologies classified into different categories. These are the most unambiguous result of the current quasar data sets.

\textit{Model comparison.---} The values of DIC for all models are reported in Table 1. According to the number of model parameters, these six cosmological models can be divided into two classes: the two-parameter models including the flat $\Lambda$CDM, flat PEDE and flat DGP; and the three-parameter models including non-flat $\Lambda$CDM, non-flat PEDE and non-flat DGP. For the two-parameter models, the flat PEDE provides the smallest information criterion result (DIC=390.68) among all of the flat cosmological models. However, we note that the difference of DIC between the flat PEDE and flat $\Lambda$CDM model is only $\Delta$DIC=0.65, which means these two models are comparable to each other according to this criterion. The flat DGP model is the worst one to explain the current lensed and unlensed quasar observations, since the DIC value it yields is the largest among the two-parameter models ($\Delta$DIC=2.99). As for the three-parameter models, the DIC results show that the non-flat $\Lambda$CDM model is still not the best one. The non-flat PEDE model, which is a little bit better than non-flat $\Lambda$CDM, performs best in explaining the quasar data (DIC=393.45), with positive evidence against non-flat DGP ($\Delta$DIC$=1.82$).

We also provide a graphical representation of the DIC results in Fig.~4 which directly shows the results in the IC test for each model. Out of all the candidate models, it is obvious that the flat PEDE and flat $\Lambda$CDM are the two most favored models in the data combination of lensed+unlensed quasars, Following them are the flat DGP, non-flat $\Lambda$CDM, and non-flat PEDE that give comparable fits to the data. According to the DIC results, the most disfavored model is non-flat DGP, with strong evidence against such cosmological scenario among the six models we study here.

\section{Conclusions} \label{sec:conclusion}

In this paper, we analyze six spatially flat and non-flat cosmological models ($\Lambda$CDM, PEDE and DGP) using a newly compiled sample of ultra-compact structure in radio quasars and strong gravitational lensing systems with quasars acting as background sources. This study is strongly motivated
by the need for revisiting the Hubble constant, spatial curvature and dark energy dynamics in the framework of different cosmological models of interests, and searching for implications for the non-flat Universe and extensions of the standard cosmological model (the spatially flat $\Lambda$CDM). The inclusion of such a newly compiled quasar sample in the cosmological analysis is crucial to this aim as it will extend the Hubble diagram to a high-redshift range, in which predictions from different cosmological models can be distinguished \citep{Capozziello}. From the constraints derived using the updated observations of quasars, we can identify some relatively model-independent features.

In all cosmological models, the cosmological parameters obtained from distinct quasar samples are consistent and the combination of the latest observations of quasars, i.e., the time-delay measurements of six strong lensing systems, four angular diameter distances to the lenses and 120 intermediate-luminosity quasars calibrated as standard rulers, would break the degeneracy between the Hubble constant and other cosmological parameters. The lensed quasar ($6D_{\Delta t}$ and $4 D_d$) and unlensed radio quasar (QSO) data combination produces the most reliable constraints. In particular, for most of cosmological model we study (the flat $\Lambda$CDM, non-flat $\Lambda$CDM, flat PEDE, and non-flat PEDE models), the derived matter density parameter is completely consistent with $\Omega_m\sim 0.30$ in all the data sets, as expected by the latest cosmological observations  \citep{Wong:2019kwg,Cao:2017ivt,LiuAPJ19,Li17,Cao20MN,Lian21} and Planck Collaboration results ($\Omega_{m}=0.3103 \pm0.0057$) \citep{Aghanim:2018eyx}. Nevertheless, the DGP model in both flat and non-flat cases shows a deviation from this prediction, with statistical lower values of $\Omega_m=0.246^{+ 0.043}_{-0.037}$ and $\Omega_m=0.253^{+0.086}_{-0.083}$ for the combined sample of 6$D_{\Delta t}$+4$D_d$+QSO. A joint analysis of the quasar sample provides model-independent estimations of the Hubble constant $H_0$, which is strongly consistent with that derived from the local distance ladder by SH0ES collaboration \citep{Riess:2019cxk} in the $\Lambda$CDM and PEDE model. However, in the framework of a DGP cosmology (especially for the flat universe), the measured value of $H_0$ with 1$\sigma$ uncertainty, which is 3.6$\sigma$ lower than the statistical estimates of the SH0ES results, demonstrates a perfect agreement with that derived from the recent Plank CMB observations \citep{Aghanim:2018eyx}. Our findings also confirm the flatness of our universe  \citep{Collett:2019hrr,Wei:2020suh,Qi2020arx}, i.e., the most unambiguous result of the current lensed and unlensed quasar observations, although there is some room for a little spatial curvature energy density in the non-flat $\Lambda$CDM, non-flat PEDE and non-flat DGP cases. Finally, we statistically evaluate which model is more consistent with the observational quasar data. Concerning the ranking of competing dark energy models, the flat PEDE is the most favored model out of all the candidate models, while the non-flat DGP is substantially penalized by the DIC criteria. However, our analysis still does not rule out dark energy being a cosmological constant and non-flat spatial hypersurfaces.

Considering the possible controversy around the systematics of the observed angular sizes of compact radio quasars \citep{Kellermann93}, the other reasonable strategy to quantify the such effect of systematics is taking $\sigma_{sys}$ as an additional free parameter, which should be fitted simultaneously with the cosmological parameters of interest. Such strategy has been extensively applied in the derivation of quasar distances based on the non-linear relation between their UV and X-ray fluxes, based on the largest quasar sample consists of 12,000 objects with both X-ray and UV observations \citep{Risaliti18}. In this paper, we perform a sensitivity
analysis by introducing an overall 5\% and 20\% systematical uncertainty to the angular size measurements of compact radio quasars, in order to investigate how the cosmological constraints on flat PEDE is altered by different choices of systematics. In the framework of such error strategy in the construction of posterior likelihood $\mathcal{L}_{QSO}$, the matter density parameter and Hubble constant respectively change to $\Omega_m=0.300^{+0.040}_{-0.034}$, $H_0=74.82^{+1.81}_{-2.04}$ $\Mpc$ (5\% systematical uncertainty), and $\Omega_m=0.331^{+0.048}_{-0.040}$, $H_0=74.93^{+1.83}_{-2.05}$ $\Mpc$ (20\% systematical uncertainty). The comparison of the resulting constraints on $\Omega_m$ and $H_0$ based on different systematical uncertainties is shown in Fig.~5. In general, one can easily check that the  derived value of $\Omega_m$ is more sensitive to the systematical uncertainties of angular size measurements, i.e., a larger systematical uncertainties will shift the matter density parameter to a relatively higher value. This illustrates the importance of a larger quasar sample from future VLBI observations based on better uv-coverage \citep{Pushkarev2015}. As a final remark, from the observational point of view, one can see that the 120 unlensed radio quasars have perfect coverage of source redshifts in six strong lensing systems ($z\sim3$). Given the usefulness of compact radio quasars and strongly lensed quasars acting as standard rulers at high redshifts, we pin our hopes on a large amount of intermediate-luminosity radio quasars detected by future VLBI surveys at different frequencies \citep{Pushkarev2015}, and strongly lensed quasars with well-measured time delays discovered by future surveys of Large Synoptic Survey Telescope  \citep{Collett2015}.

\begin{acknowledgements}

This work was supported by the National Natural Science Foundation of China under Grants Nos. 12203009, 12021003, 11690023, 11920101003 and A2020205002; the Strategic Priority Research Program of the Chinese Academy of Sciences, Grant No. XDB23000000; the Interdiscipline Research Funds of Beijing Normal University; the China Manned Space Project (Nos. CMS-CSST-2021-B01 and CMS-CSST-2021-A01); and the CAS Project for Young Scientists in Basic Research under Grant No. YSBR-006. X. Li was supported by National Natural Science Foundation of China under Grants No. 1200300, Hebei NSF under Grant No. A2020205002 and the fund of Hebei Normal University under Grants No. L2020B02.

\end{acknowledgements}

{}

\end{document}